# Hybrid-Density Functional Theory Study on Band Structures of Tetradymite-$Bi_2Te_3$, $Sb_2Te_3$, $Bi_2Se_3$, and $Sb_2Se_3$ Thermoelectric Materials

**Sudong Park and Byungki Ryu** [*]

*Thermoelectric Conversion Research Center, Korea Electrotechnology Research Institute (**KERI**),*

*Changwon 51543, Republic of Korea*

The low energy band structure near the band gap determines the electrical performance of thermoelectric materials. Here, by using the hybrid-density functional theory (hybrid-DFT) calculations, we calculate the low energy band structure of $Bi_2Te_3$, $Sb_2Te_3$, $Bi_2Se_3$ and $Sb_2Se_3$ in tetradymite phase. We find that the band structure characteristics are very sensitive the selection of the exchange energy functional. The predictability of the band gaps and band degeneracies is not enhanced in hybrid-DFT calculations, as compared to DFT calculations. The poor prediction of low energy band structures is originated from the poor prediction of interlayer distances and the high structure sensitivity on the band gap. We conclude that the hybrid DFT calculations are not superior to DFT calculations when predicting band structures of tetradymite $Bi_2Te_3$, $Sb_2Te_3$, $Bi_2Se_3$ and $Sb_2Se_3$ thermoelectric materials.

PACS numbers: 71.15.Mb, 71.20.-b, 71.20.Nr, 72.15.Jf

Keywords: Band structure, $Bi_2Te_3$, $Sb_2Se_3$, $Bi_2Se_3$, hybrid-DFT

# I. INTRODUCTION

Bi$_2$Te$_3$-related alloys, such as (Bi$_{1-x}$Sb$_x$)$_2$Te$_3$ and Bi2(Te$_{1-y}$Se$_y$)$_3$, are well known thermoelectric materials. Bi$_2$Te$_3$-related alloys have the large Seebeck coefficient and Power Factor (PF) and the small lattice thermal conductivity (κ) from high lattice anhamonicity, and thereby they have the large thermoelectric figure of merit ZT about 1 near room temperature.[1,2]

The large Seebeck coefficient of Bi$_2$Te$_3$-related materials at room temperature is originated from the high band degeneracy (*g*) of band edge states.[1,2,3] The **g** values of Bi$_2$Te$_3$ are experimentally measured to be 6 for both valence band maximum (VBM) and the conduction band minimum (CBM),[1,2] and density functional theory (DFT) calculations also yield the same band degeneracy.[3,4,5,6,7,8]

The energy band gap ($E_{gap}$) determines the high temperature thermoelectric properties. In a single parabola band model with a constant relaxation time approximation, the Seebeck coefficient (α) is a monotonic increasing function of temperature. Meanwhile the thermal conductivity (κ) without electrical thermal conductivity is a monotonic decreasing function of temperature due to the enhanced three phonon scattering. However, in real materials, the band gap is finite and the minority carriers are generated by thermal excitation across the band gap. The absolute value of Seebeck coefficient is decreased and the electric thermal conductivity is increased when temperature increases.[1] As a result, the ZT value becomes smaller when temperature exceeds certain value, where ZT is defined as (α$^2$/ρκ)T, where ρ and T are electric resistivity and absolute temperature, respectively.

The experimental band gap of Bi$_2$Te$_3$ is known to be about 0.17 eV, only a few times of room temperature thermal energy, and thereby the maximum ZT is occurred limited and smaller than 2. For PbTe, the experimental band gap is between 0.3 and 0.4 eV and the maximum ZT occurs at relatively high temperature, as compared to that of Bi$_2$Te$_3$.

It is well known that the band gap is underestimated in DFT calculations. For example, the Si DFT

band gap is 0.6 eV in DFT calculations, smaller by 0.5 eV compared to experimental value.[10] For $Bi_2Te_3$, the calculated band gap is also smaller than the experimental band gap.[3] Recent report on beyond DFT calculations reveals that the predicted band gap from the quasi particle GW approximation is consistent to the experimental band gap.[11,12] However the computational cost of band gap prediction is very challenging especially for $Bi_2Te_3$ or similar materials, because their band edges are not located at the special k-points. [12]

The hybrid-DFT calculations have been successfully adopted to predict the band gaps and the defect properties in semiconductors, insulators, and their interfaces.[13,14] The computational cost of hybrid-DFT calculations is known to be relatively smaller than that of GW calculations. However, still there are disadvantage in use of hybrid-DFT such as describing shallow defect nature and overestimation of band offsets.[15,16] Unfortunately, in our knowledge, neither the band gap or band degeneracy of $Bi_2Te_3$, $Sb_2Te_3$, and $Bi_2Se_3$ has been well studied.

In this work, we report hybrid-DFT band structures of $Bi_2Te_3$, $Sb_2Te_3$, $Bi_2Se_3$, and $Sb_2Se_3$. We validate the predictability of low energy band structure and find that the predictability of band gap and band degeneracy from the hybrid-DFT is not superior to that of the DFT calculations.

## II. CALCULATION METHOD

First-principles density functional calculations are performed to investigate the band structures of $Bi_2Te_3$, $Sb_2Te_3$, $Bi_2Se_3$, and $Sb_2Se_3$ thermoelectric materials. The Vienna-Ab-initio-Simulation-Package (VASP) [17,18] code is used for DFT and hybrid-DFT calculations with the projector augmented wave (PAW) pseudopotentials.[19] The DFT band structures are calculated using the generalized-gradient-approximation (GGA) exchange-correlation energy functional parameterized by Perdue, Burke, and Ernzerhof (PBE).[20] For hybrid-DFT band structures, we use the exchange-correlation energy functional HSE06 with a mixing parameter α of 25% and a screening parameter of 0.2 Å$^{-1}$.[21]

We use the tetradymite $Bi_2Te_3$-like structure with the experimental lattice parameter and theoretically optimized internal coordinates for all binary $Bi_2Te_3$, $Sb_2Te_3$, $Bi_2Se_3$, and $Sb_2Se_3$ compound.[3] The atomic structure is relaxed with gamma-centered $12\times12\times12$ k-point mesh. For the band structure calculations, we first perform the self-consistent-field (SCF) calculations with coarse $6\times6\times6$ k-point mesh to obtain the total charge density and then perform the non-SCF calculations with fine $12\times12\times12$ k-point mesh. Note that the band gap of $Bi_2Te_3$ is sensitive to the selection of k-point mesh.[3] Here, as the atomic elements are heavy, all band gaps are calculated with an inclusion of the spin orbit interaction (SOI).

## III. RESULTS AND DISCUSSION

The primitive unit cell of tetradymite structure $M_2Q_3$ (M=Bi or Sb, Q=Te or Se) is rhombohedral (RHL) with a space group of No.166. The atomic layers are stacked along $[111]_{RHL}$ direction, similar to FCC stacking ( ABCABC …). Two M layers, one Q(1) layers, and two Q(2) layers consist one quintuple layer (QL) such as QL = {Q(1)-M-Q(2)-Bi-Q(1)}. Each atom is surrounded by 6 neighboring atoms, with a nearly octahedral symmetry. M is surrounded by three Q(1) and three Q(2) atoms, Q(2) is surrounded by six M atoms, and Q(1) is surrounded by three M and three Q(1) in the adjacent QL. The interaction between QLs is known to be the van der Waals interaction. Each QL layer is also stacked as ABC sequence. Thereby, three are three QL with 5 M, 10 Q(1), and 5 Q(2) atoms in the hexagonal conventional unit cell. As a comparison we also consider the $Sb_2Se_3$ in the RHL phase, although its ground state is orthorhombic. The structure we used is summarized in Table. 1. Here, we use the experimental lattice parameter with the PBE relaxed internal coordinates for all species. In addition we also consider the experimental coordinates for $Bi_2Se_3$ and the HSE06 relaxed structure for $Bi_2Te_3$.

We calculate the hybrid-DFT band gap ($E_{gap}^{HSE06}$) of tetradymite-$Bi_2Te_3$, $Sb_2Te_3$, $Bi_2Se_3$, and $Sb_2Se_3$, using HSE06 functional with the inclusion of SOI. Here we adopt the experimental lattice parameters. As shown in Table 1, neither PBE nor HSE06 predict the correct band gaps. The predictability of band gap is not enhanced by using hybrid-DFT. We calculate the relative band gap error (errEg), defined as the difference between calculation and experiment $E_{gap}$ over the $E_{gap}^{expt}$. In PBE calculations with PBE relaxed geometries, the errEg values of $Bi_2Te_3$, $Sb_2Te_3$, and $Bi_2Se_3$ are −37, −24, and −13 %, respectively. In HSE calculations with PBE relaxed geometries, they are 19, 8, −88 %. Although the band gap predictability seems to be enhanced for $Bi_2Te_3$ and $Sb_2Te_3$, it is severely degraded for $Bi_2Se_3$. We also check the direct band gap and find that the band gap change is not constant in the same materials, implying that the band energy variation depends on the k-point. We would like to emphasize that the band gap difference of about 0.1 eV is very critical for thermoelectric properties prediction, because the thermoelectric materials and devices are always exposed to heat source and thereby the minority carrier transport occurs.

We investigate the effect of internal coordinates on the band gaps, especially for $Bi_2Se_3$. Both in PBE and HSE06 calculations, the band gaps are very sensitive to the internal coordinates, *u* and *v*. Here, we consider the three different geometries: PBE relaxed structure, Nakajima structure, and Wyckoff structure. Although the lattice volume changes negligibly less than 0.23 %, the distance between atomic layers significantly varies about 0.2 Å among various geometries Our PBE relaxed $Bi_2Se_3$ structure is found to be very similar to the Nakajima structure. However, the Wyckoff structure has smaller QL thickness and larger QL gap than the Nakajima structure. This structural difference affect the band structure of $Bi_2Se_3$. In PBE calculations, the band gaps of $Bi_2Se_3$ are 0.260, 0.315, and 0.227 eV for PBE relaxed, Wyckoff, and Nakajima Bi2Se3 structures, respectively. In HSE06 calculations, the band gaps change to 0.037, 0.165, and 0.092 eV. Note that the band gap from PBE is more consistent to the experimental value of ~0.3 eV. Surprisingly, the gaps from the hybrid DFT are even smaller than the PBE band gaps.

We also calculate the band gap of $Bi_2Te_3$ with HSE06 relaxed structure. In HSE06 relaxed structure, the bonding between atoms in a single QL is increased and the QL gap is reduced by 0.118 Å. In PBE calculations, the indirect band gap is decreased from 0.107 eV for PBE-relaxed structure to 0.059 eV for HSE06-relaxed structure. However, in HSE06 calculations, the indirect band gap is increased from 0.059 eV for PBE-relaxed structure to 0.167 eV for HSE06-relaxed structure. Thus, the band gap is not only sensitive to the selection of exchange correlation functional but also sensitive to the atomic structure.

Like this, the predictability of band gap is not enhanced even though we use the hybrid-DFT calculations. Sometimes, the band gap of hybrid-DFT is smaller than that of DFT. We think that this abnormal prediction behavior is strongly related to the band gap correction by spin-orbit-interaction. When the spin-orbit-interaction is included, the band edge states are mixed or inversed. Note that $Bi_2Te_3$ and $Bi_2Se_3$ are well known topological insulator (TI) material and the orbital inversion is the well-known property. In calculations, the SOI strength might be related to the energy difference, here it is the band gap. Thus, the band gap correction by SOI might be proportional to the inverse of the band gap before SOI inclusion. If non-corrected PBE band gap without SOI is smaller than the non-corrected HSE06 band gap without SOI, the band gap correction will be larger for PBE. It means that the final corrected band gap can be larger for PBE than for HSE06 due to the larger SOI correction.

Next, we investigate the band degeneracies of band edge states and their positions. The PBE calculations well predict the experimentally measured the $Bi_2Te_3$ band degeneracies of VBM and CBM ($g_{VBM}$ and $g_{CBM}$), as the quasi-particle GW calculations do.[6] In PBE calculations, the $k$ point of $Bi_2Te_3$ band edge states is at (0.583, 0.583, 0.583). In GW calculations, it is calculated to be (0.58, 0.58, 0.68).[6] In hybrid calculations, the VBM is also located at the (0.583, 0.583, 0.583) with $g_{VBM}$=6. However, the CBM state is located at the (0.167, 0.167, 0.167) with $g_{CBM}$=2, inconsistent to our PBE calculations. Thus, the hybrid calculations do not predict the experimental $g$ values, especially for the VBM of $Sb_2Te_3$ and $Bi_2Se_3$. The band degeneracy predictability of PBE calculations is superior to HSE06 calculations.

The band degeneracies are also sensitive to the internal atomic coordinates. When the interlayer distances varying by about 0.2 Å, the k point positions also varies as well as the band gap varies. In PBE calculations, $k_{VBM}$ of PBE relaxed structure is at (0.333, 0.333, 0.250) and that of Wyckoff structure is at (0.417, 0.417, 0.417).

We also perform the hybrid-DFT calculations with a different faction of exact exchange (α) and extrapolate the optimal α values for $Bi_2Te_3$, $Sb_2Te_3$, and $Bi_2Se_3$. For $Bi_2Te_3$, the indirect band gap is reduced toward the experimental value (0.17 eV), from 0.202 eV for α = 25% and 0.186 eV for α = 20%. However, the direct band gap degeneracy is still largely different from the measured value. From the band gap extrapolation from the results of α = 20% and 25 %, we expect that α = ~15% can predict the reliable band gap of $Bi_2Te_3$, as compared to conventional HSE06 functional. In the case of $Sb_2Te_3$, the extrapolated optimal α for band gap is expected to be about ~15%. However, for $Bi_2Se_3$, it seems that no mixing parameter of hybrid calculations well reproduces the experimental band gap.

## IV. CONCLUSION

In conclusion, we investigate the characteristics of low energy band structures for tetradymite-$Bi_2Te_3$, Sb2Te3, $Bi_2Se_3$, and $Sb_2Se_3$ binary compounds. The low energy electronic structure characteristics, such as band gap and band degeneracy, are very sensitive to the selection of the exchange energy functional and the internal atomic coordinates. We conclude that the hybrid DFT calculations is not superior to DFT calculations for predicting band structure of tetradymite $Bi_2Te_3$, $Sb_2Te_3$, $Bi_2Se_3$ and $Sb_2Se_3$ thermoelectric materials.

## ACKNOWLEDGMENTS

This work was supported by the National Research Foundation of Korea (NRF) grant funded by the Korea government (MSIP) (No. 2011-0030040).

**Table 1**. Atomic and electronic structure table of $Bi_2Te_3$, $Sb_2Te_3$, $Bi_2Se_3$, and $Sb_2Se_3$ compounds in tetradymite phase. $\|a_{RHL}\|$ is the length of the rhombohedral lattice vector and $\cos\theta$ is the directional cosine between the rhombohderal lattice vectors. $u$ and $v$ are internal coordinates for Bi and Te(1), where Te(2) is located at the middle of the QL. The interlayer distance is also calculated. The band gaps from experimental values and our calculated values are shown. When calculating the band gap, the SOI interaction is included. For $Bi_2Se_3$, various atomic structures are considered, including PBE relaxed structure, Wyckoff structure, and Nakajima structure. The band gap of $Bi_2Te_3$ is also calculated by using HSE06 relaxed structure. Note that the band gaps are very sensitive to the selection of exchange correlation energy functional and the internal coordiantes. However, we can not conclude that the hybrid-DFT is superior to DFT when calculating the band structures of these materials.

| | | # | 1 | 2 | 3 | 4 | 5 | 6 | 7 |
|---|---|---|---|---|---|---|---|---|---|
| | | Material | $Bi_2Te_3$ | $Sb_2Te_3$ | $Bi_2Se_3$ | $Sb_2Se_3$ | $Bi_2Se_3$ | $Bi_2Se_3$ | $Bi_2Te_3$ |
| Atomic Structure | Lattice Parameter | Relaxation | Expt. | Expt | Expt | Expt | Expt | Expt | Expt |
| | | Volume (Å$^3$) | 169.11 | 158.51 | 141.56 | 132.17 | 141.57 | 141.89 | 169.11 |
| | | $\|a_{RHL}\|$ | 10.473 | 10.426 | 9.841 | 9.794 | 9.841 | 9.840 | 10.473 |
| | | $\cos\theta$ | 0.912 | 0.917 | 0.912 | 0.916 | 0.912 | 0.911 | 0.912 |
| | | $a_{hex}$ (Å) | 4.383 | 4.250 | 4.138 | 4.004 | 4.138 | 4.143 | 4.383 |
| | | $c_{hex}$ (Å) | 30.487 | 30.400 | 28.640 | 28.553 | 28.640 | 28.636 | 30.487 |
| | Internal Parameter | Relaxation | PBE relaxed | PBE relaxed | PBE relaxed | PBE relaxed | Wyckoff position | Nakajima position | HSE06 relaxed |
| | | $u$ | 0.40000 | 0.39867 | 0.40066 | 0.39923 | 0.39900 | 0.40080 | 0.39948 |
| | | $v$ | 0.20962 | 0.21090 | 0.21090 | 0.21226 | 0.20600 | 0.21170 | 0.21156 |
| | Layer-Layer distances (Å) | Te(2)-Bi, $(u-1/3) \times c_{hex}$ | 2.033 | 1.986 | 1.928 | 1.881 | 1.881 | 1.932 | 2.017 |
| | | Bi-Te(1), $(2/3-u-v) \times c_{hex}$ | 1.739 | 1.736 | 1.578 | 1.575 | 1.766 | 1.551 | 1.696 |
| | | QL thick, $(2/3-2v) \times c_{hex}$ | 7.543 | 7.444 | 7.013 | 6.914 | 7.294 | 6.966 | 7.425 |
| | | QL gap, $(2v-1/3) \times c_{hex}$ | 2.619 | 2.689 | 2.534 | 2.604 | 2.253 | 2.579 | 2.737 |
| $E_{gap}$ (eV) | | Expt. Gap | 0.170 | 0.170 | 0.300 | | 0.300 | 0.300 | 0.170 |
| | Indirect Gap | $E_{gap}^{PBE}$ | 0.107 | 0.129 | 0.260 | 0.266 | 0.315 | 0.227 | 0.059 |
| | | $E_{gap}^{HSE06}$ | 0.202 | 0.184 | 0.037 | 0.010 | 0.164 | 0.092 | 0.059 |
| | Direct Gap | $E_{gap}^{PBE}$ | 0.107 | 0.157 | 0.336 | 0.266 | 0.442 | 0.293 | 0.167 |
| | | $E_{gap}^{HSE06}$ | 0.311 | 0.157 | 0.037 | 0.010 | 0.165 | 0.092 | 0.261 |

**Table 2**. Band degeneracy of $Bi_2Te_3$, $Sb_2Te_3$, $Bi_2Se_3$, and $Sb_2Se_3$ binary compounds are calculated for the k point of VBM, CBM, and that of direct of band gap. The band degeneracies are sensitive to the selection of exchange correlation energy functional.

| Material | $E_{XC}$ | VBM | | CBM | | Direct band gap | |
|---|---|---|---|---|---|---|---|
| | | $g_{VBM}$ | $k_{VBM}$ | $g_{CBM}$ | $k_{CBM}$ | $g_{direct}$ | $k_{direct}$ |
| $Bi_2Te_3$ | PBE | 6 | (5/12 5/12 1/3) | 6 | (5/12 5/12 1/3) | 6 | (5/12 5/12 1/3) |
| | HSE06 | 6 | (5/12 5/12 1/3) | 2 | (1/6 1/6 1/6) | 6 | (5/12 5/12 1/3) |
| $Sb_2Te_3$ | PBE | 6 | (5/12 5/12 1/3) | 2 | (1/6 1/6 1/6) | 6 | (5/12 5/12 1/3) |
| | HSE06 | 1 | (0 0 0) | 2 | (1/6 1/6 1/6) | 2 | (1/12 1/12 1/12) |
| $Bi_2Se_3$ | PBE | 6 | (1/3 1/3 1/4) | 1 | (0 0 0) | 1 | (0 0 0) |
| | HSE06 | 1 | (0 0 0) | 1 | (0 0 0) | 1 | (0 0 0) |
| $Sb_2Se_3$ | PBE | 1 | (0 0 0) | 1 | (0 0 0) | 1 | (0 0 0) |
| | HSE06 | 1 | (0 0 0) | 1 | (0 0 0) | 1 | (0 0 0) |
| $Bi_2Se_3$ | PBE | 6 | (5/12 5/12 5/12) | 1 | (0 0 0) | 6 | (1/2 1/4 1/4) |
| | HSE06 | 1 | (0 0 0) | 1 | (0 0 0) | 1 | (0 0 0) |
| $Bi_2Se_3$ | PBE | 6 | (1/3 1/3 1/4) | 1 | (0 0 0) | 1 | (0 0 0) |
| | HSE06 | 1 | (0 0 0) | 1 | (0 0 0) | 1 | (0 0 0) |
| $Bi_2Te_3$ | PBE | 6 | (5/12 5/12 1/3) | 6 | (5/12 5/12 1/3) | 6 | (5/12 5/12 1/3) |
| | HSE06 | 6 | (5/12 5/12 1/3) | 2 | (1/12 1/12 1/12) | 2 | (1/6 1/6 1/6) |